# Scaling of losses with size and wavelength in nanoplasmonics and metamaterials.

Jacob B Khurgin
*Johns Hopkins University, Baltimore MD USA*
Greg Sun
*University of Massachusetts, Boston MA, USA*

We show that, for the resonant metal-dielectric structures with sub-wavelength confinement of light in all three dimensions, the loss cannot be reduced significantly below the loss of the metal itself unless one operates in the far IR and THz regions of the spectrum or below. Such high losses cannot be compensated by introducing gain due to Purcell-induced shortening of recombination times. The only way low loss optical meta-materials can be engineered is with as yet unknown low loss materials with negative permittivity.

Recent years have seen significant progress in two interrelated fields of nanoplasmonics (NP) and metamaterials (MM) [1-3]. Both of these research directions rely upon the most remarkable feature of sub-wavelength metallic objects – a high degree of concentration of electro-magnetic fields achievable in their vicinity which is well beyond the concentration allowed by the diffraction limit. In turn, the ability to concentrate energy near these "artificial atoms" allows one to arrange them in a regular manner and thus engender new metamaterials with optical properties that are unattainable in natural material, the elusive negative refractive index [3] being just one of them. While significant steps in developing functional NP and MM devices have been made, widespread practical implementation of them has been impeded by many factors, the most significant of which remains the inherent loss associated with the absorption in the metal. It is well known that the rate of energy loss in the metal, determined mostly by electron-phonon and to a lesser degree electron-electron scattering inside the conduction band is on the order of $2\gamma \sim 10^{14} s^{-1}$ for noble metals and it gets even larger at shorter $\lambda$'s where the band-to-band absorption arises. In the Drude approximation one can represent the dielectric constant of a metal as $\varepsilon(\omega) = \varepsilon_r - j\varepsilon_i = 1 - \omega_p^2 / (\omega^2 + j\omega\gamma)$ where $\omega_p$ is plasma velocity and $\gamma$ is the velocity relaxation rate, which is exactly one half of the energy loss rate. The quality factor of metal $Q_m = |\varepsilon_r|/\varepsilon_i \approx \omega/\gamma$ is less than 40 even under most optimistic projections and in reality far worse than that once surface scattering has been factored in. $Q_m$ in turn determines the $Q$ of the whole device and thus its ability to perform its requisite function, e.g. achieve high field concentration, enhance radiative decay via Purcell's effect, or provide a sharp resonance capable of reversing the sign of the magnetic permeability. While significant efforts are being devoted to the search for alternative negative materials, such as highly doped semiconductors [4] or more exotic ones [5] as well as to compensating the loss with gain [6], such efforts have not yet yielded practical results, and most of the researchers are trying to avoid, or at least mitigate, the metal losses by means of design, *i.e.* trying to find the configurations and spectral regions where the losses are effectively reduced. While reduced total losses for different material combinations and $\lambda$'s have been widely reported, these results are not systematic and often are no more than the consequence of moving to longer $\lambda$'s, where interband absorption can be avoided. Overall the most impressive MM results have been obtained in the THz region of the spectrum and beyond [7], while results in the visible and especially UV have been far more modest. This problem has been first tackled in the pivotal works [8,9] where the difficulty in scaling to the short λ's is explained by the rise of "kinetic inductance" associated with the inertia of electrons in small nanostructures saturating the resonance frequency. In layman's terms, there are simply not enough electrons in the metal to achieve resonances at the short λ's. But these studies explain only one side of the scaling difficulties associated with the inertia of electrons – no systematic study of scaling limitations associated with electron-related loss has been made. In this work we perform a simple study of the Q-factor of two highly representative NP and MM elements – a split ring resonator (SRR) and an elliptical nanoparticle (ENP) – as a function of their dimensions and λ and demonstrate that the apparent improvement of Q in the mid IR is purely material-related (avoidance of interband absorption) while the improvement in the THz region and beyond is of a more fundamental nature (*i.e.*, present in Drude metals) and is associated with an increase in the conductivity current relative to the displacement current. The point of this study is to show that high Q-factors in optical/plasmonic regime $\omega \gg \gamma$ are fundamentally unattainable with existing metals.

A simple general rationalization of this statement can be given before we embark on the more involved derivations using the Maxwell equation

$$\nabla \times \boldsymbol{H} = j\omega\varepsilon\varepsilon_0 \boldsymbol{E} = j\omega\varepsilon_r\varepsilon_0 \boldsymbol{E} + \sigma \boldsymbol{E} \qquad (1)$$

where $\sigma = \omega\varepsilon_i\varepsilon_0 = \varepsilon_0\omega_p^2/\gamma$ is the conductivity of the medium. Now, if we assume that the scale on which the electric field is confined is roughly *a*, integrating (1) over the area $\pi a^2$ and invoking Stokes theorem yields this relation for the field magnitudes

$$|H| \sim \frac{a}{2}|j\omega\varepsilon\varepsilon_0 + \sigma||E| = \frac{a\omega}{2}\sqrt{\varepsilon_r^2 + \varepsilon_i^2}\,\varepsilon_0|E| \qquad (2)$$

Now, when the frequency is close to resonance, E and H (with current J) are oscillating 90 degrees out of phase with each other. Hence half of the time all the energy is stored in electro-static form, $U_E = \varepsilon_0^2 \varepsilon_r^2 E^2 V / 2$ while the other half of the time the energy is expected to be stored in the magnetic form $U_M = \mu^2 \mu_0^2 H^2 V / 2$. Their ratio is

$$\frac{U_M}{U_E} \sim \frac{\mu_0 \mu \varepsilon_0^2 (\varepsilon_r^2 + \varepsilon_i^2)}{\varepsilon_0 \varepsilon_r} \left(\frac{a\omega}{2}\right)^2 \sim \left(\frac{\pi n a}{\lambda}\right)^2 \left(1 + \frac{\varepsilon_i^2}{\varepsilon_r^2}\right) \sim \left(\frac{\pi n a}{\lambda}\right)^2 \left(1 + \frac{\gamma^2}{\omega^2}\right) \quad (3)$$

where $n$ is the refractive index. It is easy to see from here that, as long as $\varepsilon_i \ll \varepsilon_r (\omega \gg \gamma)$ and the confinement scale $a \ll \lambda/n$, the magnetic energy is but a tiny fraction of the electro-static energy (the well-known electro-static limit) and, from an energy conservation point of view, it follows that half of the time practically all the energy is stored in the kinetic motion of electrons where it is inevitably lost at the rate $2\gamma$. Hence for truly sub-$\lambda$ confinement the rate of energy loss near resonance is always close to $\gamma$ as is indeed confirmed by calculations for spherical nanoparticles near surface plasmon (SP) resonance [10]. The electrostatic limit, however, is no longer valid at lower frequencies $\omega \sim \gamma$ and less, as conduction rather than displacement current becomes dominant in (1) and H is no longer inversely proportional to $\lambda$. Thus, a smaller fraction of energy is contained in the kinetic motion of electrons, and, rather ironically, as the $Q_m$ decreases below unity, the Q of the whole device increases!

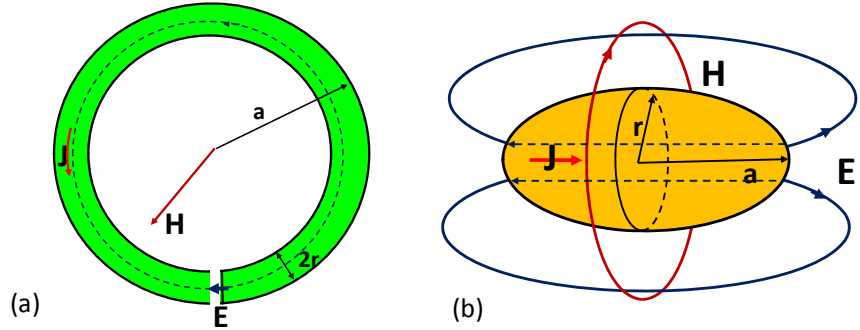

**Fig. 1** Geometries and fields for (a) split ring resonator and (b) elliptical nanoparticles

To confirm and quantify these predictions, we first consider an example of an SRR shown in Fig. 1a that consists of a ring of radius $a$ made of a cylindrical wire with radius $r$ and one or more gap capacitors $C$ that are treated as adjustable to provide a required resonant frequency $\omega \sim 1/\sqrt{LC}$. When current flows through the wire energy is stored in two forms – magnetic energy $U_M = L_M I^2 / 2$, where $L_M \approx \mu_0 a \ln(8a/r)$ is the conventional (magnetic) inductance and the kinetic energy of the electrons $U_K = L_K I^2 / 2$, where the kinetic inductance is $L_K = 2\pi a / A\varepsilon_0 \omega_p^2$ [8], and where $A$, the area over which the current flows subject to skin effect, can be found as

$$A(\omega) \approx \pi r^2 \left(1 + \frac{\pi r}{\lambda_p \sqrt{1 + 2/Q_m(\omega)}}\right)^{-1} \quad (4)$$

Now, as mentioned above, at resonance the total energy alternates between potential (electro-static energy in the capacitor) and kinetic, i.e. $U_M + U_K$. $U_K$ dissipates with rate $2\gamma$=and, since the energy spends only one half of the time in the kinetic form, the dissipation rate of the whole structure is $\gamma_{SRR} = \gamma L_K / (L_K + L_M)$. The $\lambda$-depended Q is then

$$Q_{SRR}(\lambda_\gamma) = \frac{1}{\lambda_\gamma \gamma'} + \frac{2\pi^2}{\gamma'} \ln \frac{8}{\beta} \beta a_\lambda Q_{m0} \left[\frac{1}{\beta a_\lambda \lambda_\gamma Q_{m0}} + \frac{\pi}{\sqrt{1 + 2\lambda_\gamma \gamma'}}\right]^{-1} \quad (5)$$

where $\beta = r/a$ is the diameter/thickness ratio, $a_\lambda = a/\lambda$ is the "sub-wavelength" parameter of the structure, $Q_{m0} = \omega_p / \gamma$ is the maximum $Q$ of metal in the Drude model (of course in real metal this $Q$ is never achieved since near plasma frequency losses increase dramatically – it is used just as a parameter here), and $\lambda_\gamma = \lambda / \lambda_b = Q_m^{-1}(\lambda)$ is the ratio of $\lambda$ to the "border wavelength" $\lambda_b = 2\pi c / \gamma$ at which $\omega = \gamma$, i.e. what we can loosely define as a border between electronics and optics. Parameters for the Drude model of gold [11] ($\omega_p$=13.06×10$^{15}$ s$^{-1}$, $\gamma$=12.3×10$^{13}$ s$^{-1}$) yield $Q_{m0} = 110$ and $\lambda_b$~15 μm. Finally, the factor $\gamma' = \gamma(\omega)/\gamma$ accounts for the discrepancy between the actual metal permittivity [11] and the Drude value.

In Figs. 2a, and 2b, the results for Q$_{SRR}$ and the ratio of the effective loss rate $\gamma_{eff}(\lambda) = \omega / Q_{SRR}(\lambda)$ to the Drude value of the loss in metal are shown as functions of wavelength for a rather thick wire $\beta$=0.25 and four different values of $a_\lambda$. The dashed lines are obtained using the Drude approximation while the solid lines represent results obtained using experimental values of gold permittivity [11]. As one can see, the initial sharp increase in Q$_{SRR}$ as $\lambda$ moves from visible towards near IR is associated solely with the intrinsic reduction of metal loss as photon energy becomes insufficient to cause the band-to-band transitions in the metal. But once the metal loss settles at a constant Drude value, the improvement in $Q_{SRR}$ becomes less dramatic for larger SRR's ($2a=\lambda/4$) and is reversed for smaller, truly sub-$\lambda$ SRR's. The $Q_{SRR}$ starts recovering only in the vicinity of the "border wavelength" and from there increases roughly as $\lambda^{1/2}$ as the very last term in (5) becomes dominant. The behavior of $\gamma_{eff}(\lambda)$ plotted in Fig. 2b is similar – the initial rapid decrease is followed by an essentially flat region when $\gamma_{eff}(\lambda) \approx \gamma$, exactly as predicted above, and only when wavelength exceeds $\lambda_b$, i.e., when one is essentially in electronic rather than optical domain, the loss decreases as roughly $\lambda^{-3/2}$. Notice that for true sub-wavelength SRR with a diameter less than $\lambda/8$ the reasonably low loss rate of 1/100ps cannot be achieved until $\lambda \sim 300 \mu m$, and for a loss rate of 1/ns, one would have to venture well into the mm wave domain!

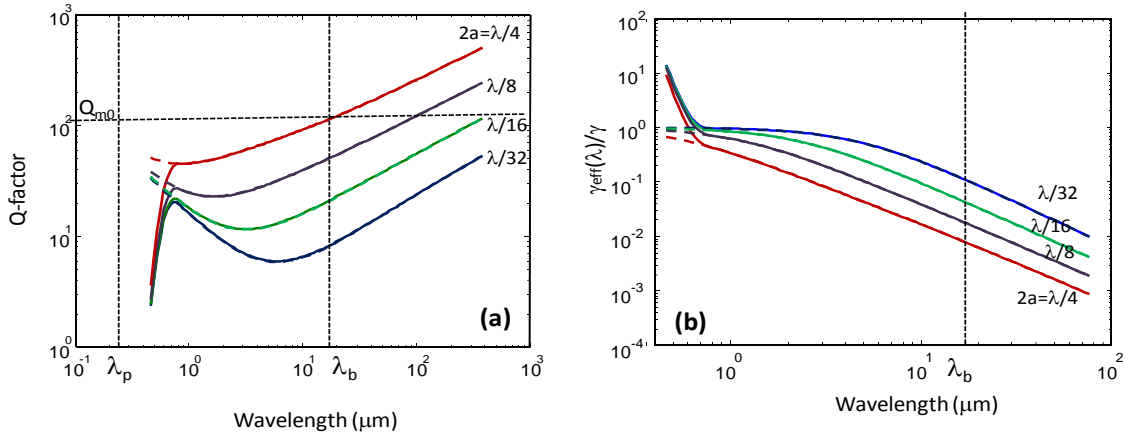

**Fig. 2** Q-factor (a) and effective loss rate (b) in the SRR with r=a/4 and different diameters 2a. Solid lines: exact values of ε, dashed lines: Drude model

Next we consider the losses in the prolate ENP (Fig.1b), used extensively for local field enhancement [1], with large half-axis *a* and two equal small half-axes *r*. Unlike the SRR where ω can be tuned over a wide range by adjusting the gap(s), in the ENP the resonant frequency is determined only by the eccentricity. Thus for a given value of *a, r* and *l* are related as roughly $a/r \approx \lambda^2/3\varepsilon_D\lambda_p^2 + (\varepsilon_D - 1)/3\varepsilon_D$ [12] where $\varepsilon_D$ is the dielectric constant of the material in which the ENP is embedded. For long λ's we can then approximate $\beta \approx 3\varepsilon_D / \lambda_\gamma^2 Q_{mo}^2$ and

$$Q_{ENP} = \frac{1}{\lambda_\gamma \gamma'} + \frac{12 n_D \pi^3}{\gamma' \lambda_\gamma^2 Q_{m0}} \ln\left(\sqrt{\frac{2}{3}} \frac{\lambda_\gamma Q_{m0}}{n_D}\right) a_\lambda \left[\frac{\lambda_\gamma Q_{m0}}{3 n_D a_\lambda} + \frac{\pi}{\sqrt{1+2\lambda_\gamma \gamma'}}\right]^{-1} \tag{6}$$

where $n_D = \varepsilon_D^{1/2}$ and the length $a$ is now normalized to the wavelength in the dielectric.

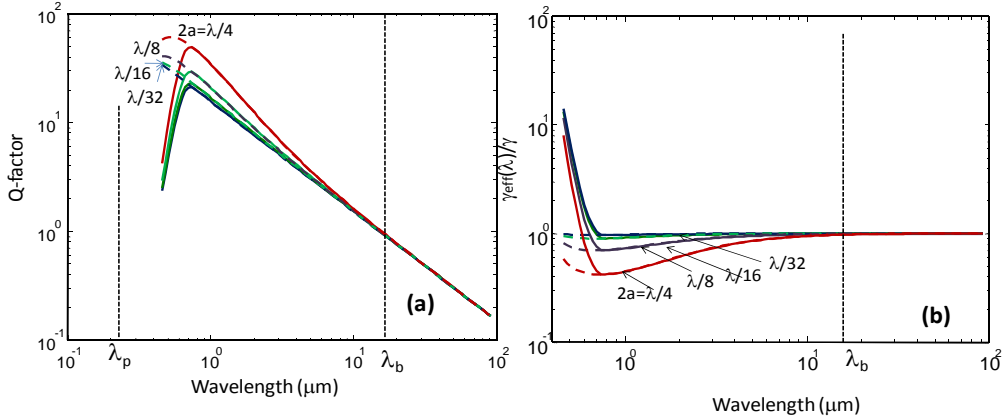

**Fig. 3** Q-factor (a) and effective loss rate (b) in the ENP of different lengths 2a. Solid lines: exact values of ε, dashed lines: Drude model

As Fig 3 shows for all values of *2a/λ* the effective loss rate after the initial dive quickly settles at exactly the value of half of the metal loss and the Q factor accordingly falls as $\lambda^{-1}$. This is simply the consequence of the fact that in order to shift resonance to a long λ one has to reduce the ellipsoid cross-section to very low values and thus increase kinetic inductance. For field enhancement, clearly ENP's work best at the short λ's that are just long enough to avoid the band-to-band absorption in the metal, e.g. close to 1 μm in gold. At these λ's their $Q_{ENP}$ can be as large as or a bit larger than $Q_{SRR}$, which makes ENP ideally suited for sensing. At longer λ's it is preferable to use SRR or some structures in between, for example nano-antennas comprised of ENP's with gaps between them to provide adjustable capacitance to tune the resonance into the red. Nevertheless, it is fair to say that at long wavelengths their Q won't exceed that of an SRR. Hence the key conclusion of our work remains valid and states that in true sub-wavelength metal-dielectric structures the loss cannot be reduced significantly below the one half of the metal loss for as long as the frequency is larger than metal scattering rate. Only when metal resistance exceeds the kinetic inductance, i.e. one essentially operates in classical electronics rather than plasmonics regime the losses are reduced as a larger and larger fraction of energy is stored in the magnetic field.

One can extend these results to propagating SP polaritons (SPP) on the metal-dielectric boundaries, where the degree of confinement in the direction normal to the interface is the effective width while in the interface plane the degree of confinement is the polariton's wavelength that near SP resonance becomes indeed very short. Unfortunately, once that wavelength becomes just a few times shorter than wavelength in dielectric, the loss will inevitably become equal to precisely one half of the metal loss, i.e. $\gamma$ [13]. Once again, it is essential to state that our results are applicable only to the case of **sub-wavelength confinement in all three dimensions (3D)** –the moment the mode becomes comparable to half-wavelength in just one direction, as is the case of nanowires [14], MSM structures [15], slot [16], or whispering gallery mode plasmons [17,18], H-field becomes large enough to suck the energy out of the kinetic motion of electrons and the loss is greatly reduced, in fact to the point where it can be compensated by the gain in the dielectric and lasing can be attained in small volumes, less than $(\lambda/n)^3$ [14,15,18]. Yet if one examines the reports of sub-$\lambda$ lasing, one cannot avoid noticing that, in at least in one dimension, the mode size always remains larger than $(\lambda/n)$.

Although SP lasing, or "spasing" [19,20] is not the main subject of this work, one can still make a simple order-of magnitude estimate of what it would take to compensate the projected modal loss $\gamma_{eff} \sim 10^{14} s^{-1}$ in a true sub-$\lambda$ mode. In a semiconductor gain medium, obviously a very high injection density, of the order of $10^{19}$-$10^{20}$cm$^{-3}$ depending on $\lambda$ would be required. This density is high, but not-unattainable provided the spontaneous recombination rate (RR) is reasonably slow. But this, regrettably, goes against the physics of sub-$\lambda$ cavities where spontaneous RR is greatly enhanced by the Purcell effect [21], (along with the stimulated emission rate). Hence the injection rate (current) required to maintain carrier density becomes prohibitively high. In fact, simple rate equation for the number of SPP's in the mode (similar to the photon equations in the laser [22]), $\dot{n}_{SPP} = g_{eff}(n_{SPP}+1) - \gamma_{eff} n_{SPP}$, shows that when modal gain $g_{eff}$ is close to compensating the loss, the spontaneous RR approaches $\gamma_{eff} \sim 10^{14} s^{-1}$, meaning that injection current $i_{inj} \sim e\gamma_{eff} \sim 10 \mu A$ must flow into the very small modal volume. For 50 nm diameter mode that corresponds to the threshold current density just under 1 *MA/cm²*, or at the very least 2 orders higher than in conventional semiconductor lasers [23]. Once such additional phenomena as spontaneous lifetime quenching by other modes, Auger recombination (prevalent at high carrier densities), and the increase in $\gamma$ due to surface scattering are taken into account, the actual threshold current density is likely to be on the order of 10 $MA/cm^2$ which would require electron and hole drift velocities well in excess of $10^6$ *cm/s* that have never been observed in highly doped semiconductors even in most pure samples and simple geometries unimpeded by metal structures. Thus we cannot perceive how the true sub-$\lambda$ in all 3D "spaser" [19, 20] can be feasible at optical $\lambda$'s , at least when it comes to its most practical electrically pumped implementation.

In conclusion, we have shown that when one operates with sub-$\lambda$ confinement $a \ll \lambda$ in 3D and at frequencies that are higher than the scattering rate in the metal ( $\omega \gg \gamma$ ) *i.e.*, in what is referred to as "plasmonics" rather than "optics" ( $a \geq \lambda$ ) or "electronics" ( $\omega \leq \gamma$ ) regime, no amount of clever engineering can reduce the modal loss significantly below $\gamma$ . In hindsight, this conclusion appears obvious, as in order to reap all the benefits of "plasmonics" a significant fraction of energy must constantly flow in and out of "plasma", (*i.e.* the inherently lossy motion of free electrons in the metal). Yet this fact is not commonly recognized, so we hope that our work will show the limits of engineering the shape of NP and MM structures. Furthermore, we have pointed out that since the recombination time in sub-$\lambda$ structures is shortened by the Purcell the pump rates required to compensate the loss with gain appear to be impractically high. That leaves finding a better material with negative ε the only viable alternative to bring losses in the plasmonics down. In the absence of such material plasmonics and metamaterials still can find use in selective niches such as sensing, where loss is not the key factor.

**List of figures**